\documentclass[twocolumn,10pt]{IEEEtran}
\usepackage{indentfirst}
\usepackage{amsmath}
\usepackage{amsfonts}
\usepackage{amssymb}
\usepackage{cite}
\usepackage{times}
\usepackage[dvips]{graphicx}
\usepackage{hhline}
\usepackage{color}
\usepackage{multirow}
\usepackage{epstopdf}
\usepackage{algpseudocode}
\usepackage{algorithm}
\usepackage{caption}
\usepackage{setspace}

\usepackage{graphicx}
\usepackage{adjustbox}
\usepackage{subfigure}

\usepackage{balance}

\usepackage{tikz}
\usepackage{pgfplots} 
\pgfplotsset{compat=newest} 
\pgfplotsset{plot coordinates/math parser=false} 
\newlength\figureheight 
\newlength\figurewidth 
\usepgfplotslibrary{patchplots}
\usepgfplotslibrary{fillbetween}

\newcommand{\bi}{\begin{itemize}}
\newcommand{\ei}{\end{itemize}}

\newcommand{\be}{\begin{IEEEeqnarray}}
\newcommand{\ee}{\end{IEEEeqnarray}}

\newcommand{\commentLB}[1]{}
\usepackage{multicol}
\usepackage{etoolbox}

\usepackage{geometry}
\usepackage[absolute]{textpos}
\usepackage{everyshi}

\begin{document}

\title{Classification of grasping tasks based on EEG-EMG coherence}

\author{
Giulia Cisotto, Anna V. Guglielmi, Leonardo Badia and Andrea Zanella \\
Dept.\ of Information Engineering, University of Padova, 
via Gradenigo 6B, 35131 Padova, Italy \\
email: \{cisottog, guglielm, badia, zanella\}@dei.unipd.it
}

\date{}
\maketitle
\thispagestyle{empty}
\pagestyle{empty}

\begin{abstract}
This work presents an innovative application of the well-known concept of cortico-muscular coherence for the classification of various motor tasks, i.e., grasps of different kinds of objects.
Our approach can classify objects with different weights (motor-related features) and different surface frictions (haptics-related features) with high accuracy (over $0.8$).
The outcomes presented here provide information about the synchronization existing between the brain and the muscles during specific activities; thus, this may represent a new effective way to perform activity recognition.
\end{abstract}

\begin{textblock*}{17cm}(1.7cm, 0.5cm)
\noindent\scriptsize This paper will be published on \emph{2018 IEEE $20^{th}$ International Conference on e-Health Networking, Applications and Services (Healthcom), Ostrava, September 2018.}\\
\textbf{Copyright Notice}: \textcopyright 2018 IEEE. Personal use of this material is permitted. Permission from IEEE must be obtained for all other uses, in any current or future media, including reprinting/republishing this material for advertising or promotional purposes, creating new collective works, for resale or redistribution to servers or lists, or reuse of any copyrighted component of this work in other works.
\end{textblock*}


\section{Introduction}

This paper presents a brand-new application of \emph{cortico-muscular coherence} (CMC) for the classification of various motor tasks, i.e., grasps of different kinds of objects.

CMC is a well-known concept in the neuro-rehabilitation field, defined as the coherence function between an electroencephalographic (EEG) and an electromyographic (EMG) signal, well characterized in boh healthy subjects \cite{Matsuya2017} and those affected by different kinds of motor-related diseases.

The CMC function accounts for the amount of synchronization between brain and muscular activity at each frequency and strongly depends on the particular motor task performed by the individual, e.g., precision grips (fine hand movements), stable (\emph{isometric}) contractions or rapid muscle contractions~\cite{Baker1997,Mima2000,Kilner2000}.

The aim of this contribution is to test to what extent CMC can classify motor-related features, as the weight of a grasped object, as well as haptics-related features, as its friction surface.
In addition, we study how the relevant parameters of the CMC computation (e.g., duration of segments, kernel of the classifier) impact the classification performance.
Furthermore, we evaluate the improvement of classification accuracy when data from several muscles is jointly used to distinguish different objects.

We show that CMC can reliably classify objects characterized by different weights and surfaces with high accuracy, over $0.8$.
As a consequence, the paper advocates the capability of joint EEG and EMG processing, i.e., using CMC, to robustly classify motor activities, which would be a brand-new method in the field of motion analysis and activity recognition.

The paper is organized as follows. Section~\ref{sec:methods} describes the CMC computation procedure and presents the classification problem. Section~\ref{sec:results} shows the most relevant outcomes and, finally, Section~\ref{sec:discussion} discusses the impact of this work and its possible evolutions.

\section{Methods}\label{sec:methods}

\subsection{CMC computation} \label{sec:computation}
The processing pipeline to get CMC from EEG and EMG signals consists of a simple and widely-used procedure\cite{halliday}\cite{Ushiyama2010}: each signal is segmented into \emph{trials} of few seconds ($1$ to $4$~s), then a pass-band filter is applied to cancel out frequency components that are known not to carry relevant information (typical pass-band is $3$-$80$~Hz). Signal processing is fully implemented in Matlab. All EMG segments are further (full-wave) rectified and their envelopes are extracted to get the \emph{activation profile} of the muscles\cite{konrad}.

Let $S_{x}^{(i)}(f)$ and $S_{y}^{(i)}(f)$, with $i = 1, 2, ..., N$, be the power spectra obtained by the $i-$th segment of the EEG and EMG signals, respectively. Similarly, let $S_{xy}^{(i)}(f)$ be the $i-$~th segment of the the cross-power spectrum between EEG and EMG.
Now, $S_{x}(f)$, $S_{y}(f)$ and $S_{xy}(f)$ are the averages along trials, computed from $S_{x}^{(i)}(f)$, $S_{y}^{(i)}(f)$ and $S_{xy}^{(i)}(f)$, respectively, with $i = 1, 2, ..., N$.
Similarly, the power spectra obtained by computing the standard deviation among single-trial power spectra, at each frequency, were also computed and reported in the following.
Hence, the CMC of every single trial is commonly obtained as follows:
\begin{equation}
|CMC^{(i)}(f)|^2 = \dfrac{|S_{xy}^{(i)}(f)|^2}{S_{x}^{(i)}(f)S_{y}^{(i)}(f)}.
\end{equation}

\begin{figure*}[htbp]
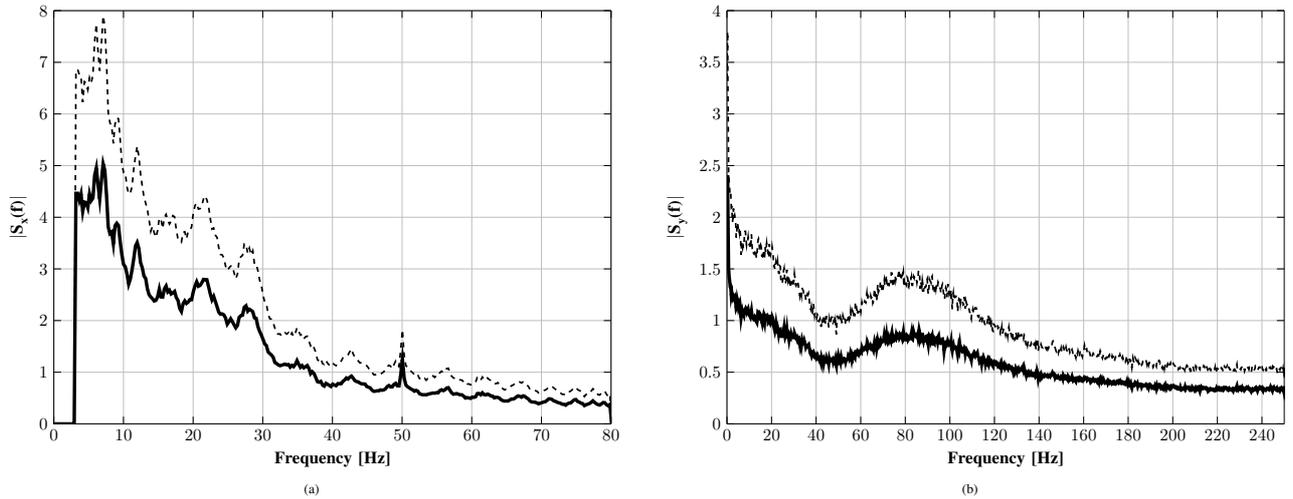

	\centering
	\resizebox {1\textwidth} {!} {
	\centering
	\captionsetup{justification=centering}
	\subfigure[]
	{\input{example_Pxx_4s.tex}}
	\hspace{5mm}
	\subfigure[]
	{\input{example_Pyy_4s.tex}}
	}
	\caption{Example of average power spectra from artefact-free segments of EEG and EMG for the \emph{light} scenario. (a) $S_{x}(f)$, (b) $S_{y}(f)$. Solid line represents the mean spectrum, dashed line the standard deviation spectrum.}\label{fig:powspectra}
\end{figure*}

As per the Cauchy-Schwarz inequality, it holds:
\begin{equation}
0 \leq |S_{xy}^{(i)}(f)|^2 \leq S_{x}^{(i)}(f)S_{y}^{(i)}(f),
\end{equation}
$|CMC^{(i)}(f)|^2$ represents the normalized cross power spectrum between EEG and EMG. Consequently, $|CMC(f)|^2$ assumes values between $0$ (uncorrelated signals) and $1$ (perfect linear relationship).
The power spectrum of every segment is computed by means of the Fast Fourier Transform (FFT) algorithm.
An example of typical $S_{x}(f)$ and $S_{y}(f)$ spectra is reported in Fig.~\ref{fig:powspectra}.

In the CMC spectrum different frequency sub-bands can be identified. Specifically, we considered the following (eight sub-bands): low-$\alpha$ ($6$-$8$~Hz), $\alpha$ ($8$-$12$~Hz), low-$\beta$ ($13$-$20$~Hz), high-$\beta$ ($20$-$30$~Hz), $\beta$ ($13$-$30$~Hz), low-$\gamma$ ($30$-$60$~Hz), high-$\gamma$ ($60$-$80$~Hz) and $\gamma$ ($30$-$80$~Hz)~\cite{Mima2000}.

\subsection{Segmentation}
Since the time required to reach, grasp, and lift the object is variable (depending on the subjective reaction time), the length of trials is variable. Therefore, EMG activation pattern can be used to precisely identify the time instant when each muscle starts to be involved in the task.
To do that, every EMG envelope is processed by a moving average filter with a $400$~ms window, i.e., high-pass filter with cut-off frequency at $2.5$~Hz.
Then, the minimum ($min$) and the maximum ($max$) values (contraction levels) are identified and a threshold is set to $TH = (max - min)/3 + min$. The period of time where the signal stays above that threshold is considered.
Finally, the half-time value of each interval ($t_0$) is used to extract the segment starting at $t_0 - DUR/2$ and ending at $t_0 + DUR/2$ (with $DUR$ either equal to $1$~s, $2$~s or $4$~s).
The corresponding EEG segment is then extracted and used to computed the CMC of that segment, as described in Section~\ref{sec:computation}.
All EEG and EMG segments have been visually inspected to remove artefacts (of any duration). The remaining artefact-free segments are then considered in the subsequent classification analysis.

\subsection{Classification and performance measurement} \label{sec:classification}
CMC is used to classify various weight or surface characteristics of the object grasped.

\begin{figure*}
 \centering
  \captionsetup{justification=centering, margin=2cm}
 {\includegraphics[width=15cm]{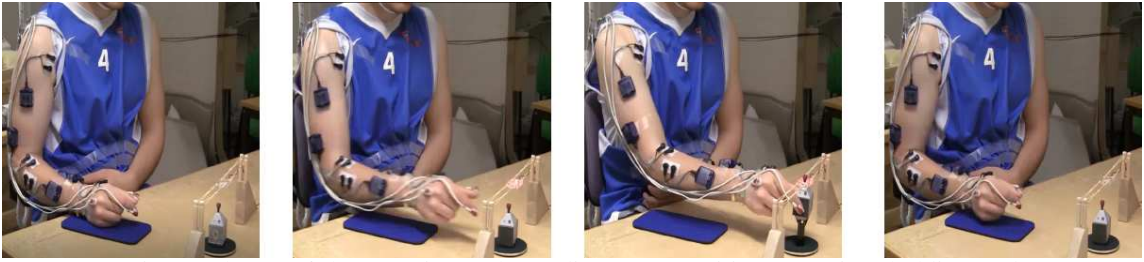}}
 \caption{Phases of the experimental paradigm: rest position, grasping, holding and releasing (modified from\cite{Luciw}).}\label{fig:paradigm}
\end{figure*}

We use the publicy-available dataset \emph{WAY-EEG-GAL} provided within the framework of the European project WAY (Wearable interfaces for hand function recovery grasp-and-lift)~\cite{Luciw}.
The dataset consists of simultaneous EEG and EMG data acquired during several repetitions of a grasp-and-lift task.
At each \emph{trial}, the participant was asked to grasp an object with their thumb and index fingers, to lift it up to a predetermined position, hold it for few seconds (see Fig.\ref{fig:paradigm}) and after that release it.

The object was unexpectedly modified in its weight (light = $165$~g, medium = $330$~g, heavy = $660$~g), surface friction (sandpaper, suede, silk) or both, according to a random pattern.
The dataset includes $84$ trials for the light condition, $57$ for the heavy condition (irrespective of the kind of surface), and $51$ where the object was covered by sandpaper and $221$ with silk (irrespective of their weight).
In particular, we denoted as \emph{heavy} the trials where the object weight was $660$~g and \emph{light} those where the weight was $165$~g.
Thirty-two EEG channels were used, with the electrodes located at standard locations (following the \emph{International 10-20 EEG System}~\cite{Nuwer1998}) on the participant scalp.
Five bipolar EMG electrodes were employed to acquire the activity from five different muscles.
Both EEG and EMG signals were downsampled to $500$~Hz sampling frequency and the effective bit resolution was $12$~bit. 
For the subsequent analysis, we considered only the data from Subject $7$ that, based on the experimental records (available online), were collected without any experimental problem.

Then, a binary linear supervised classifier, i.e., a support vector machine (SVM)~\cite{Bishop2006} is employed to distinguish among (i) light/heavy objects or (ii) sandpaper/silk surface frictions.
For the classification, we define a set of features obtained from $|CMC(f)|^2$ in each frequency band of interest (see Section \ref{sec:computation}) is used.
Finally, the performance of the classification are quantified in terms of accuracy, i.e., the ratio between the number of correct classifications (both classes of weight or surface) over the total number of instances to classify.

\section{Results} \label{sec:results}
Here we report the accuracy of the classification using different muscles and segment durations, in order to test whether, and to what extent, CMC is able to classify different grasping tasks.

The classifier kernel has been selected based on its ability to classify classes. Specifically, the linear kernel and the radial basis function (RBF) kernel~\cite{Bishop2006} are used.

\begin{figure*}[htbp]
	\centering
	\subfigure[]
	{\includegraphics[width=0.4\textwidth]{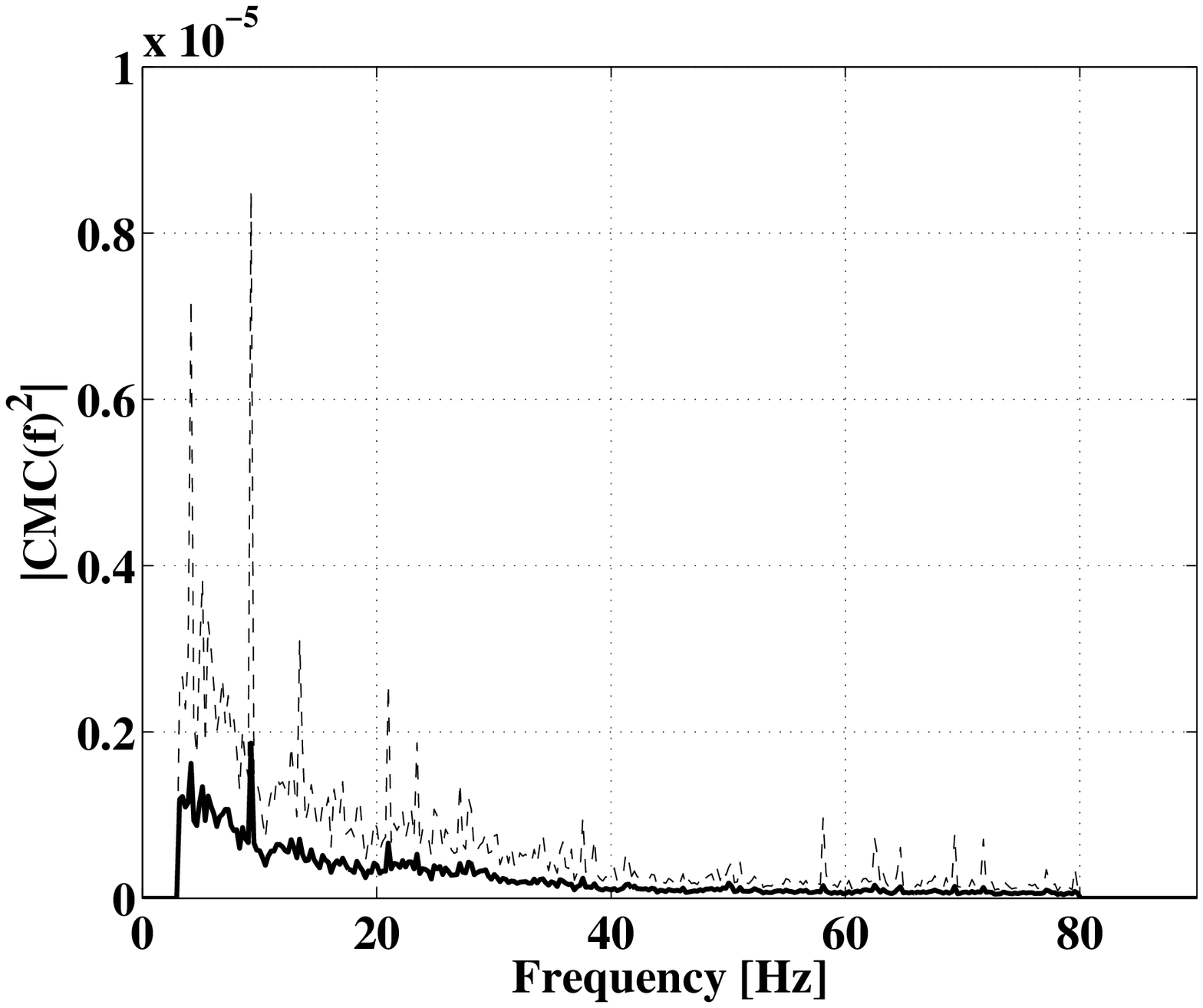}}
	\hspace{5mm}
	\subfigure[]
	{\includegraphics[width=0.4\textwidth]{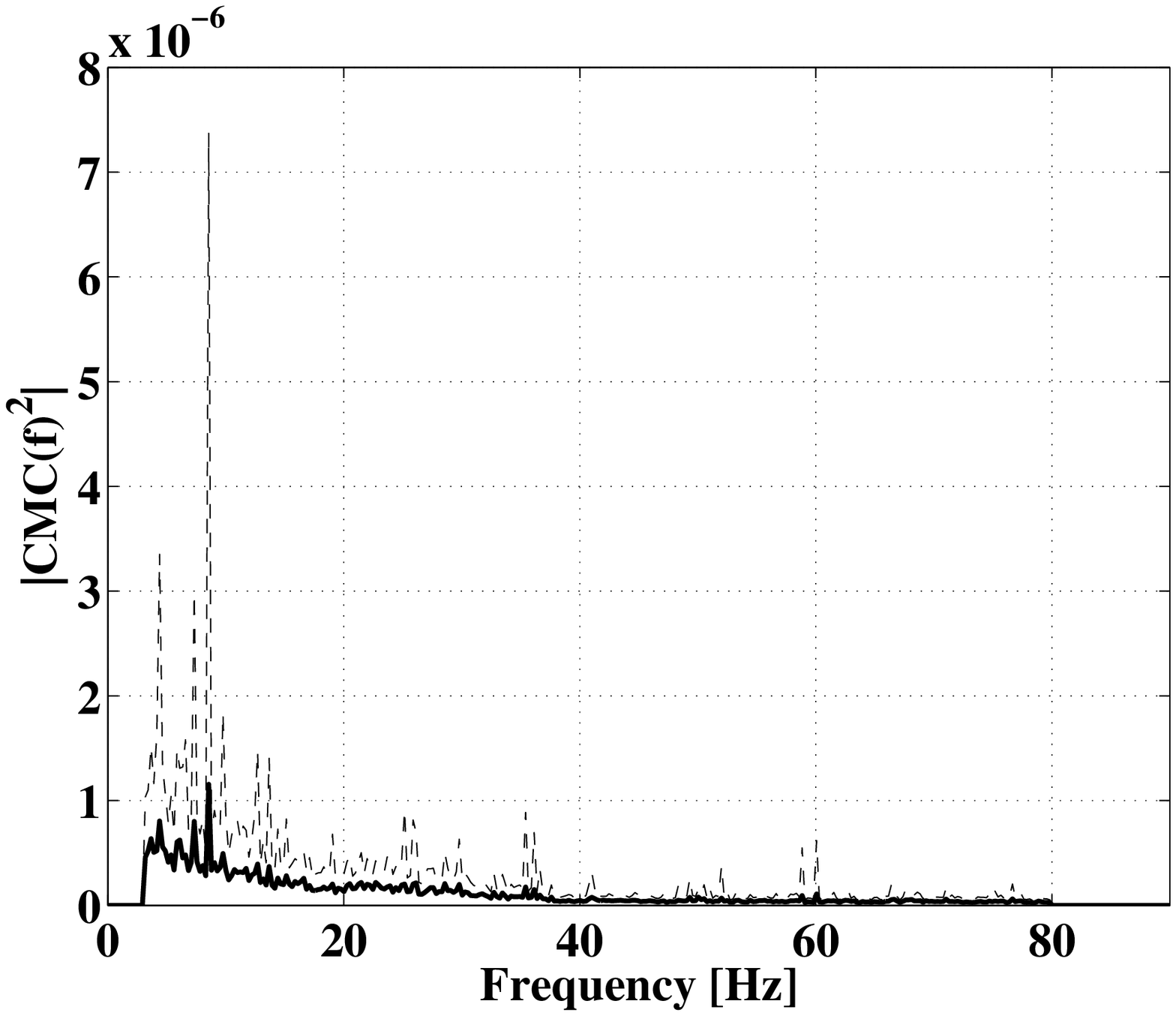}}
	\hspace{5mm}
	\subfigure[]
	{\includegraphics[width=0.4\textwidth]{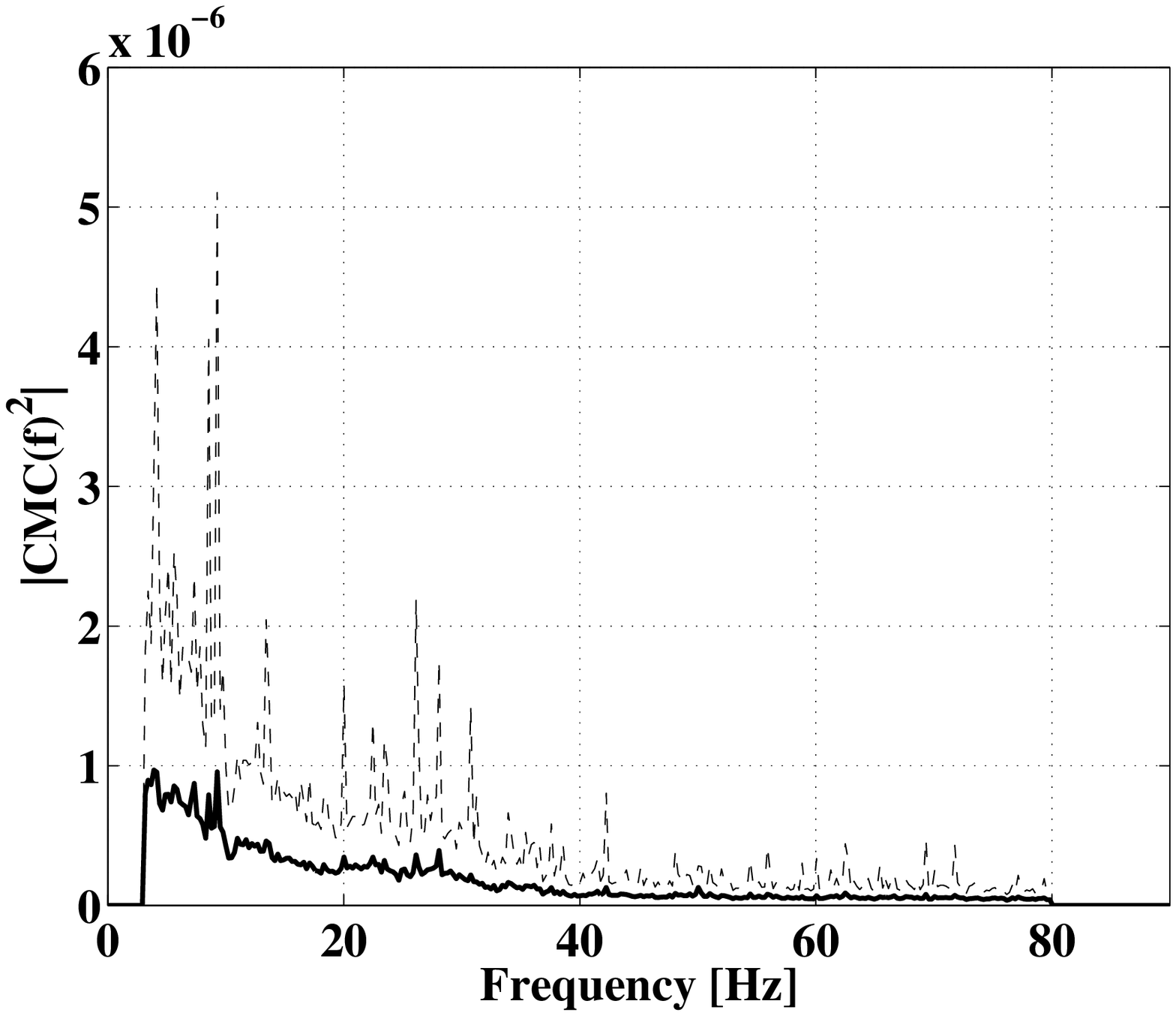}}
	\hspace{5mm}
	\subfigure[]
	{\includegraphics[width=0.4\textwidth]{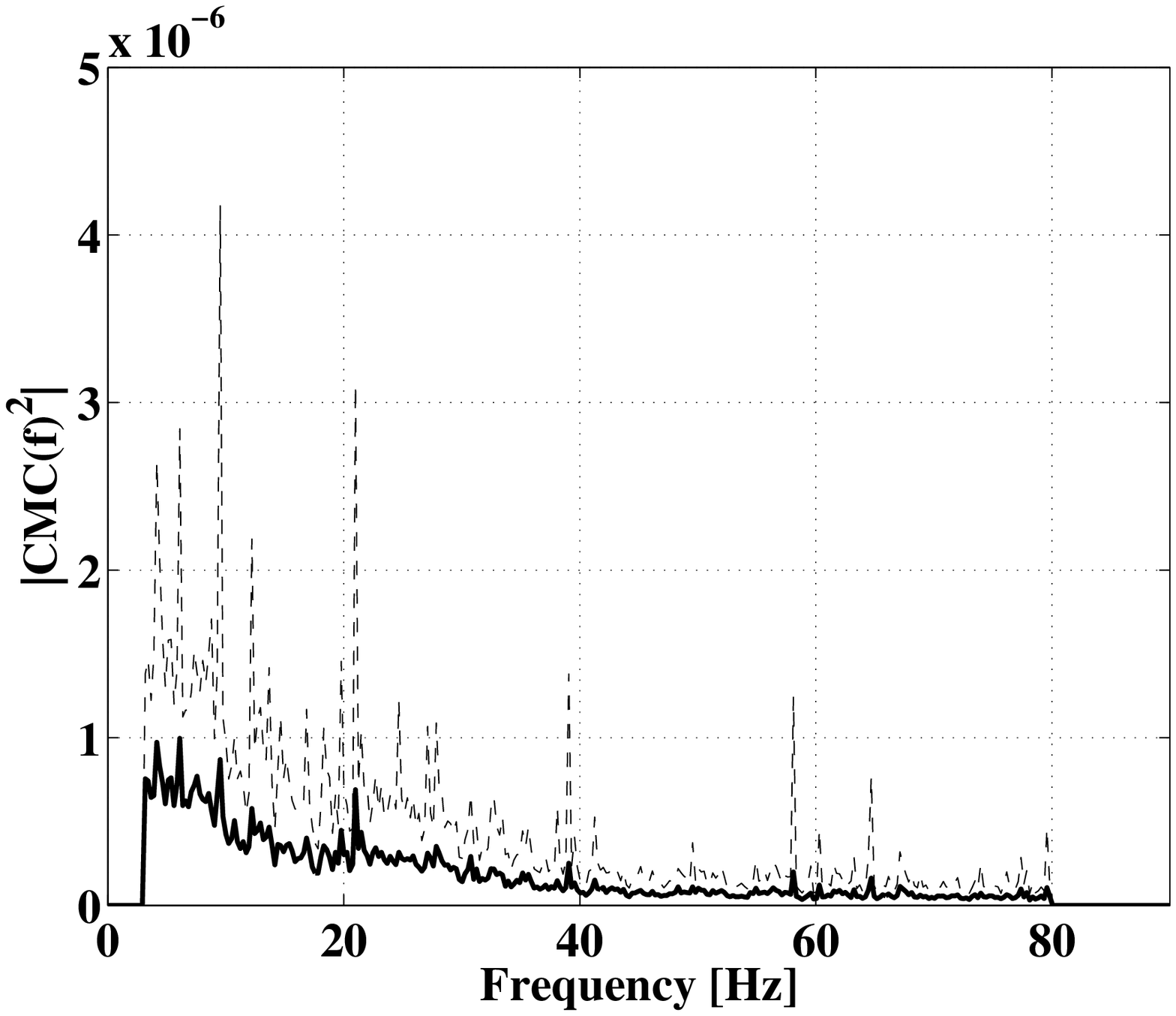}}
	\caption{Average $|CMC(f)|^2$ spectra from EEG ($C3$) and EMG ($BR$) artefact-free segments of duration $4$~s in all conditions: (a) light, (b) heavy, (c) sandpaper,  (d) silk. Solid line represents the mean spectrum, dashed line the standard deviation spectrum.}\label{fig:CMCclasses}
\end{figure*}

\begin{figure*}[htbp]
	\subfigure[]
	{\includegraphics[width=0.3\textwidth]{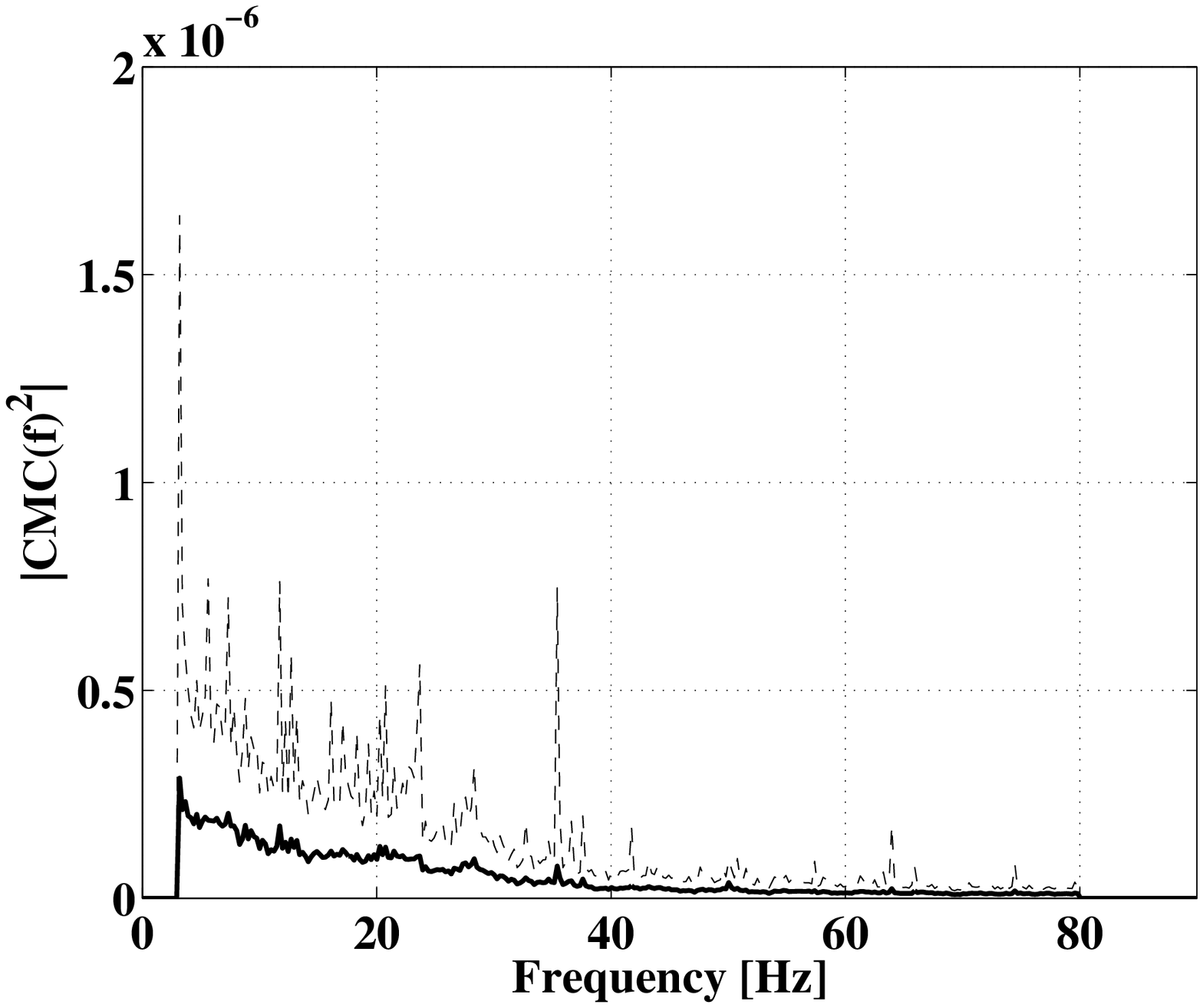}}
	\hspace{5mm}
	\subfigure[]
	{\includegraphics[width=0.3\textwidth]{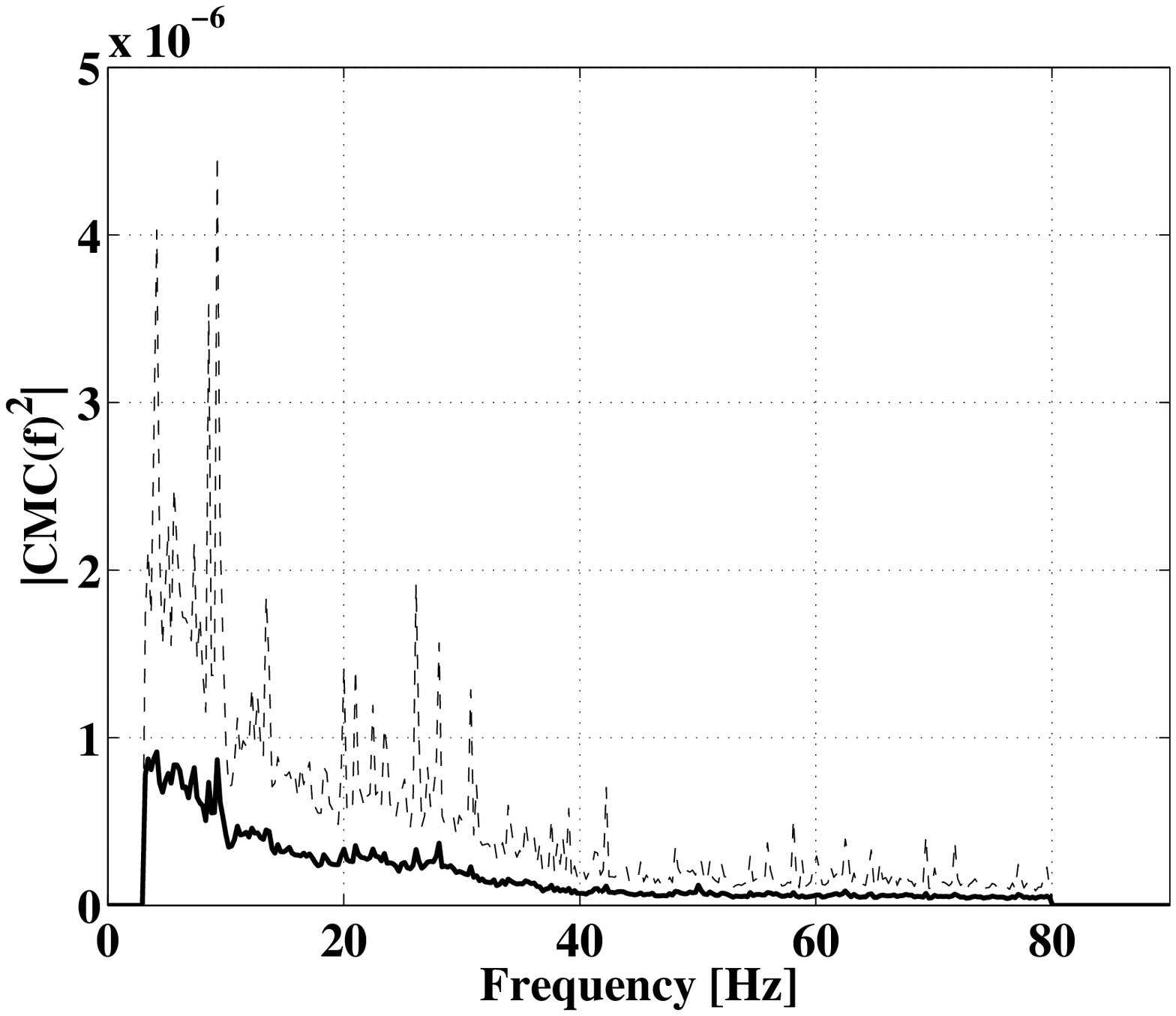}}
	\hspace{5mm}
	\subfigure[]
	{\includegraphics[width=0.3\textwidth]{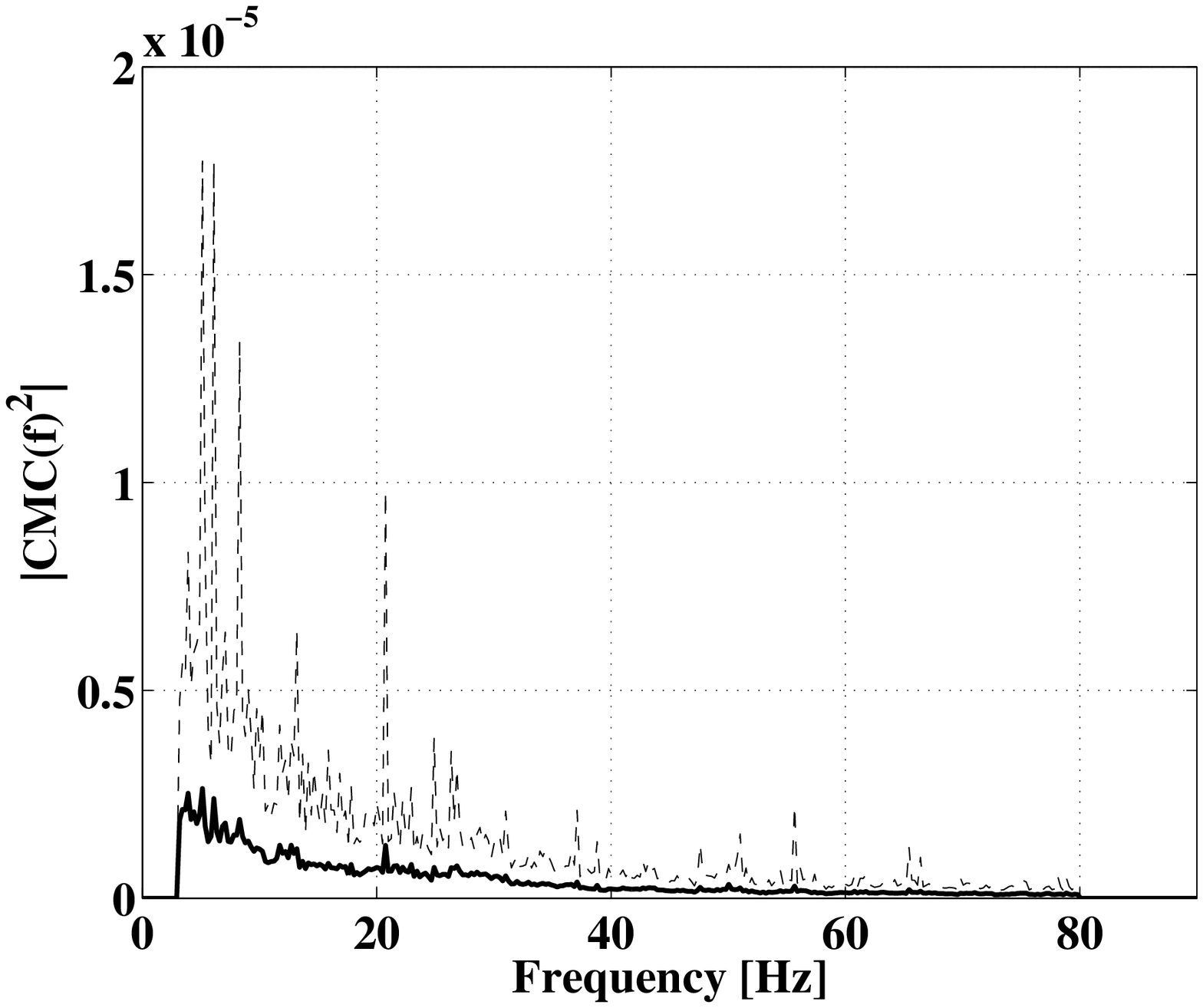}}
	\centering
	\hspace{5mm}
	\subfigure[]
	{\includegraphics[width=0.3\textwidth]{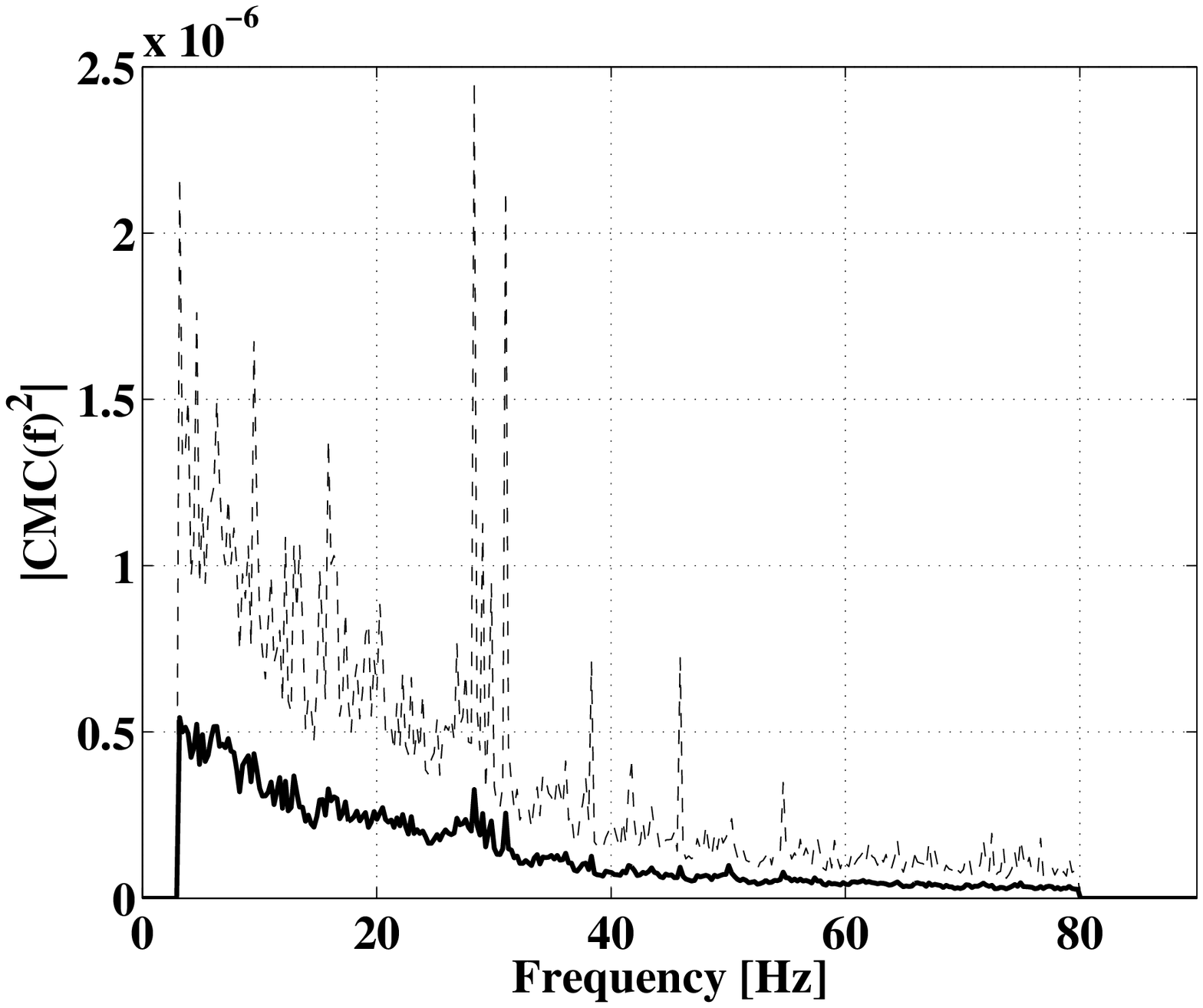}}
	\hspace{5mm}
	\subfigure[]
	{\includegraphics[width=0.3\textwidth]{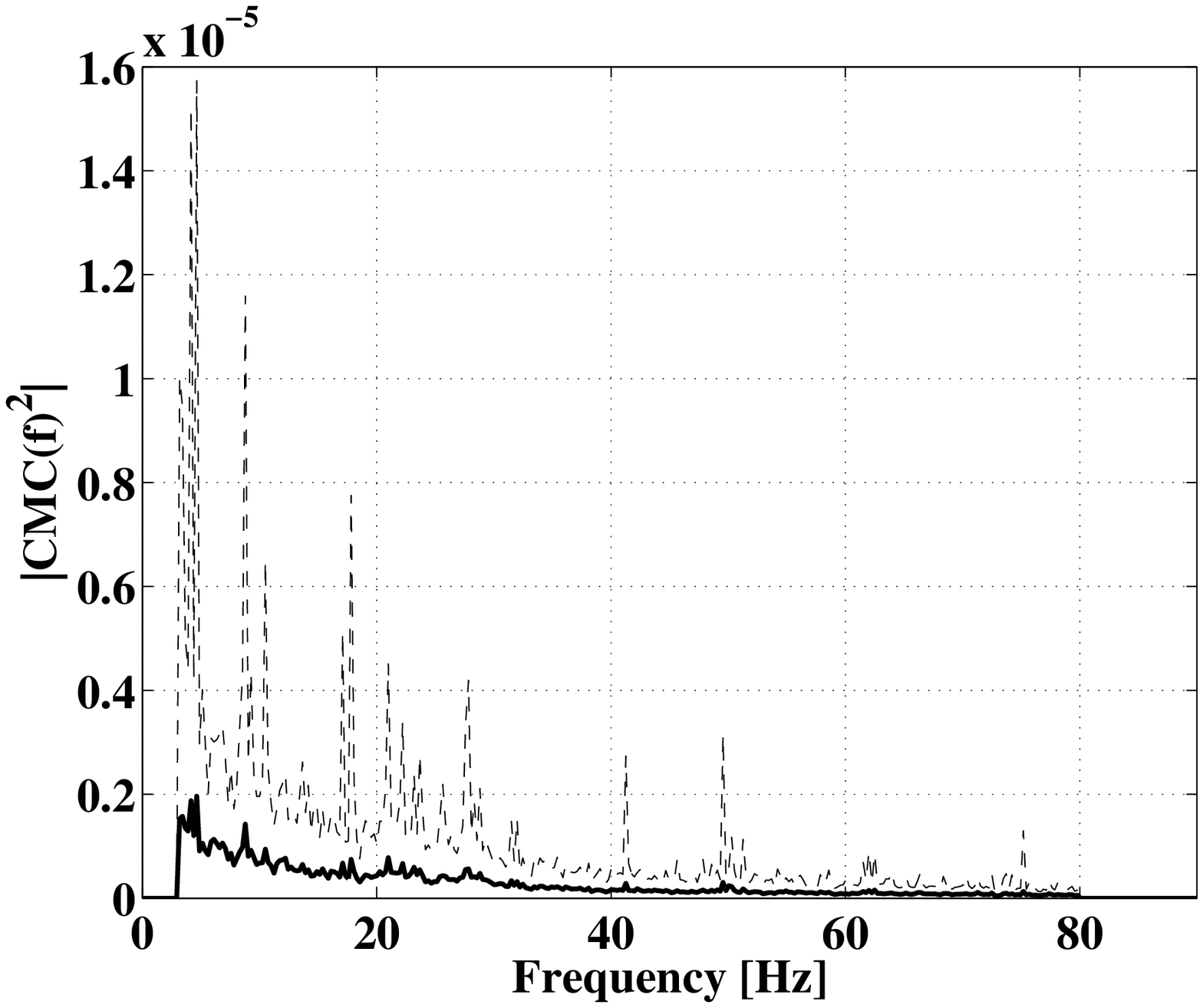}}		
	\caption{Average $|CMC(f)|^2$ spectra from artefact-free segments of duration $4$~s using $C3$ EEG electrode and all trials (irrespective for the different conditions). The EMG muscle used for CMC computation varies: (a) $AD$, (b) $BR$, (c) $FD$  (d) $CED$, (e) $FDI$. Solid line represents the mean spectrum, dashed line the standard deviation spectrum.}\label{fig:CMCmuscles}
\end{figure*}

Before that, the reliability of the preprocessed data has been tested and the differences in CMC have been evaluated by comparing CMC spectra for every class and muscle: Figs.~\ref{fig:CMCclasses} and ~\ref{fig:CMCmuscles} report the mean and standard deviation spectra in all these cases (with $C3$ selected as EEG signal and segments of duration $4$~s).
As expected, three main frequency components are visible in all spectra, i.e., at about $10$~Hz ($\alpha$ band), $20$~Hz (low-$\beta$ band) and $30$~Hz (high-$\beta$ or low-$\gamma$ band).
From a qualitative point of view, CMC spectra related to different weights are characterized by components belonging to the lower part of the spectrum, i.e., below $30$~Hz, while spectra related to different surface frictions generally show additional peaks in the upper part of the spectrum, i.e., above $30$~Hz and below $60$~Hz.


%

Moreover, Table~\ref{tab:table1} reports the results of the classification of light/heavy objects by using a single muscle with different segment lengths. 
It is worth noting that different muscles perform quite differently in separating the classes, i.e., light from heavy. For example, the most effective classification, where the highest accuracy values are reached, is - in order - with $BR$ ($0.94$), $FDI$ ($0.84$), and $FD$ ($0.82$), and segments of duration $4$~s.
A similar situation is found for sandpaper/silk surfaces (Table~\ref{tab:table2}): $AD$ ($1$~s), $FDI$ ($2$~s), and $BR$ ($4$~s) display the best accuracies ($0.69$, $0.69$ and $0.67$, respectively), therefore highlighting a different pattern of muscle activation with respect to the light/heavy case.

\begin{table}
\caption{Accuracy in the classification of light/heavy classes for different muscle/duration pairs.}\label{tab:table1}
\begin{center}

\begin{tabular}{l*{4}{c}r}
Muscle 		&\textbf{$1$~s} 	& \textbf{$2$~s} 	&\textbf{$4$~s} 	\\
\hline
AD 			& 0.66 				& 0.62 				& 0.62				\\
BR 			& 0.84 				& 0.90 				& 0.94				\\
FD 			& 0.72 				& 0.78 				& 0.82				\\
CED 		& 0.58 				& 0.66 				& 0.64				\\
FDI 		& 0.77 				& 0.79 				& 0.84				\\

\end{tabular}
\end{center}
\end{table}

\begin{table}[htbp]
\caption{Accuracy in the classification of sandpaper/silk classes for different muscle/duration pairs.}\label{tab:table2}
\begin{center}

\begin{tabular}{l*{4}{c}r}
Muscle 		&\textbf{$1$~s} 	& \textbf{$2$~s} 	&\textbf{$4$~s} 	\\
\hline
AD 			& 0.69 				& 0.63 				& 0.64				\\
BR 			& 0.57 				& 0.63 				& 0.67				\\
FD 			& 0.44 				& 0.55 				& 0.54				\\
CED 		& 0.50 				& 0.53 				& 0.54				\\
FDI 		& 0.66 				& 0.69 				& 0.54				\\

\end{tabular}
\end{center}
\end{table}    
    
\begin{figure*}[htbp]
	\centering
	\resizebox {1\textwidth}{!} {
	\centering
	\captionsetup{justification=centering}
	\subfigure[]
	{
%
\begin{tikzpicture}

\begin{axis}[%
width=4.446in,
height=3.38in,
at={(0.833in,0.667in)},
scale only axis,
area style,
separate axis lines,
every outer x axis line/.append style={black},
every x tick label/.append style={font=\color{black}},
every x tick/.append style={black},
xmin=0,
xmax=6,
xlabel style={font=\bfseries},
xlabel={No. Muscles},
every outer y axis line/.append style={black},
every y tick label/.append style={font=\color{black}},
every y tick/.append style={black},
ymin=0.5,
ymax=1.1,
ylabel style={font=\bfseries},
ylabel={Accuracy},
axis background/.style={fill=white},
xmajorgrids,
ymajorgrids,
legend style={at={(0.97,0.03)}, anchor=south east, legend cell align=left, align=left, fill=white}
]
\addplot[fill=none, draw=none, forget plot] table[row sep=crcr]{%
1	0.624456838396941\\
2	0.750654671424252\\
3	0.793159535891935\\
4	0.811162356992494\\
5	0.814814814814821\\
}
\closedcycle;
\addplot[fill=white!70!blue, draw=none, forget plot] table[row sep=crcr]{%
1	0.074197434317228\\
2	0.0434536201144589\\
3	0.021162409697618\\
4	0.00407528601502505\\
5	0\\
}
\closedcycle;
\addplot [color=blue, line width=1.0pt, forget plot]
  table[row sep=crcr]{%
1	0.624456838396941\\
2	0.750654671424252\\
3	0.793159535891935\\
4	0.811162356992494\\
5	0.814814814814821\\
};
\addplot [color=blue, line width=1.0pt, forget plot]
  table[row sep=crcr]{%
1	0.698654272714169\\
2	0.794108291538711\\
3	0.814321945589553\\
4	0.815237643007519\\
5	0.814814814814821\\
};
\addplot [color=red, line width=2.5pt, mark size=4.0pt, mark=asterisk, mark options={solid, red}]
  table[row sep=crcr]{%
1	0.707888888888889\\
2	0.808370370370371\\
3	0.816000000000003\\
4	0.814814814814821\\
5	0.814814814814821\\
};

\addplot [color=blue, line width=2.5pt, mark=o, mark options={solid, blue}]
  table[row sep=crcr]{%
1	0.661555555555555\\
2	0.772381481481482\\
3	0.803740740740744\\
4	0.813200000000006\\
5	0.814814814814821\\
};

\addplot [color=blue, dashed, line width=2.0pt, name path=B]
  table[row sep=crcr]{%
1	0.698654272714169\\
2	0.794108291538711\\
3	0.814321945589553\\
4	0.815237643007519\\
5	0.814814814814821\\
};

\addplot [color=blue, dashed, line width=2.0pt, forget plot, name path=A]
  table[row sep=crcr]{%
1	0.624456838396941\\
2	0.750654671424252\\
3	0.793159535891935\\
4	0.811162356992494\\
5	0.814814814814821\\
};

\addplot [
        thick,
        color=blue,
        fill=blue, 
        fill opacity=0.4
    ]
    fill between[
        of=A and B,
        soft clip={domain=0:1},
    ];

\end{axis}
\end{tikzpicture}%
}
	\hspace{5mm}
	\subfigure[]
	{
%
\begin{tikzpicture}

\begin{axis}[%
width=4.446in,
height=3.38in,
at={(0.833in,0.667in)},
scale only axis,
area style,
separate axis lines,
every outer x axis line/.append style={black},
every x tick label/.append style={font=\color{black}},
every x tick/.append style={black},
xmin=0,
xmax=6,
xlabel style={font=\bfseries},
xlabel={No. Muscles},
every outer y axis line/.append style={black},
every y tick label/.append style={font=\color{black}},
every y tick/.append style={black},
ymin=0.5,
ymax=1.1,
ylabel style={font=\bfseries},
ylabel={Accuracy},
axis background/.style={fill=white},
xmajorgrids,
ymajorgrids,
]
\addplot[fill=none, draw=none, forget plot] table[row sep=crcr]{%
1	0.594224998130274\\
2	0.755192297058361\\
3	0.799693801202458\\
4	0.812539301672654\\
5	0.814814814814821\\
}
\closedcycle;
\addplot[fill=white!70!blue, draw=none, forget plot] table[row sep=crcr]{%
1	0.135268522257969\\
2	0.0503561466240192\\
3	0.0258938790765699\\
4	0.00406213739544348\\
5	0\\
}
\closedcycle;
\addplot [color=blue, line width=1.0pt, forget plot]
  table[row sep=crcr]{%
1	0.594224998130274\\
2	0.755192297058361\\
3	0.799693801202458\\
4	0.812539301672654\\
5	0.814814814814821\\
};
\addplot [color=blue, line width=1.0pt, forget plot]
  table[row sep=crcr]{%
1	0.729493520388244\\
2	0.805548443682381\\
3	0.825587680279028\\
4	0.816601439068098\\
5	0.814814814814821\\
};
\addplot [color=red, line width=2.5pt, mark size=4.0pt, mark=asterisk, mark options={solid, red}]
  table[row sep=crcr]{%
1	0.742851851851852\\
2	0.816444444444446\\
3	0.845148148148148\\
4	0.817962962962966\\
5	0.814814814814821\\
};

\addplot [color=blue, line width=2.5pt, mark=o, mark options={solid, blue}]
  table[row sep=crcr]{%
1	0.661859259259259\\
2	0.780370370370371\\
3	0.812640740740743\\
4	0.814570370370376\\
5	0.814814814814821\\
};

\addplot [color=blue, dashed, line width=2.0pt, name path=B]
  table[row sep=crcr]{%
1	0.729493520388244\\
2	0.805548443682381\\
3	0.825587680279028\\
4	0.816601439068098\\
5	0.814814814814821\\
};

\addplot [color=blue, dashed, line width=2.0pt, forget plot, name path=A]
  table[row sep=crcr]{%
1	0.594224998130274\\
2	0.755192297058361\\
3	0.799693801202458\\
4	0.812539301672654\\
5	0.814814814814821\\
};

\addplot [
        thick,
        color=blue,
        fill=blue, 
        fill opacity=0.4
    ]
    fill between[
        of=A and B,
        soft clip={domain=0:1},
    ];

\end{axis}
\end{tikzpicture}%
}
	\hspace{5mm}
	\subfigure[]
	{
%
\begin{tikzpicture}

\begin{axis}[%
width=4.446in,
height=3.38in,
at={(0.833in,0.667in)},
scale only axis,
area style,
separate axis lines,
every outer x axis line/.append style={black},
every x tick label/.append style={font=\color{black}},
every x tick/.append style={black},
xmin=0,
xmax=6,
xlabel style={font=\bfseries},
xlabel={No. Muscles},
every outer y axis line/.append style={black},
every y tick label/.append style={font=\color{black}},
every y tick/.append style={black},
ymin=0.5,
ymax=1.1,
ylabel style={font=\bfseries},
ylabel={Accuracy},
axis background/.style={fill=white},
xmajorgrids,
ymajorgrids,
legend style={at={(0.97,0.03)}, anchor=south east, legend cell align=left, align=left, fill=white}
]
\addplot[fill=none, draw=none, forget plot] table[row sep=crcr]{%
1	0.663010738746492\\
2	0.750996446918275\\
3	0.803628408916412\\
4	0.813990769496489\\
5	0.814814814814821\\
}
\closedcycle;
\addplot[fill=white!70!blue, draw=none, forget plot] table[row sep=crcr]{%
1	0.0274451891736813\\
2	0.0541774765338217\\
3	0.018261700685702\\
4	0.00298142396999701\\
5	0\\
}
\closedcycle;
\addplot [color=blue, line width=1.0pt, forget plot]
  table[row sep=crcr]{%
1	0.663010738746492\\
2	0.750996446918275\\
3	0.803628408916412\\
4	0.813990769496489\\
5	0.814814814814821\\
};
\addplot [color=blue, line width=1.0pt, forget plot]
  table[row sep=crcr]{%
1	0.690455927920173\\
2	0.805173923452097\\
3	0.821890109602114\\
4	0.816972193466486\\
5	0.814814814814821\\
};
\addplot [color=red, line width=2.5pt, mark size=4.0pt, mark=asterisk, mark options={solid, red}]
  table[row sep=crcr]{%
1	0.694999999999999\\
2	0.802518518518522\\
3	0.824777777777779\\
4	0.818148148148151\\
5	0.814814814814821\\
};

\addplot [color=blue, line width=2.5pt, mark=o, mark options={solid, blue}]
  table[row sep=crcr]{%
1	0.676733333333333\\
2	0.778085185185186\\
3	0.812759259259263\\
4	0.815481481481487\\
5	0.814814814814821\\
};

\addplot [color=blue, dashed, line width=2.0pt, name path=B]
  table[row sep=crcr]{%
1	0.690455927920173\\
2	0.805173923452097\\
3	0.821890109602114\\
4	0.816972193466486\\
5	0.814814814814821\\
};

\addplot [color=blue, dashed, line width=2.0pt, forget plot, name path=A]
  table[row sep=crcr]{%
1	0.663010738746492\\
2	0.750996446918275\\
3	0.803628408916412\\
4	0.813990769496489\\
5	0.814814814814821\\
};

\addplot [
        thick,
        color=blue,
        fill=blue, 
        fill opacity=0.4
    ]
    fill between[
        of=A and B,
        soft clip={domain=0:1},
    ];

\end{axis}
\end{tikzpicture}%
}
	}
  \caption{Accuracy vs no. of muscles used for the classification of sandpaper/silk covered objects using (a) $1$~s segments, (b) $2$~s segments or (c) $4$~s segments. Mean (blue solid line) and standard deviation (blue dashed lines) of the accuracy (among all possible combinations) together with the best case (red line) are represented.}
  \label{fig:comboSS}
\end{figure*}
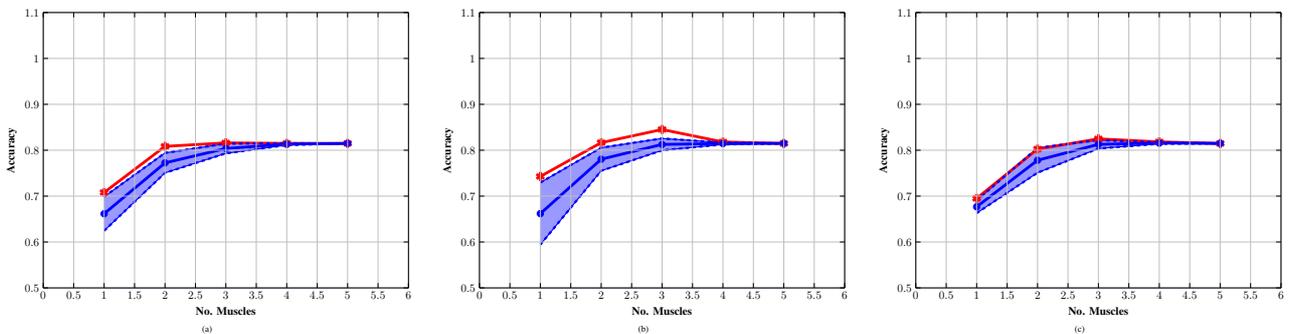

In other words, muscles that yield the highest classification accuracy can be considered as the most informative for identifying a specific task. 
If several EMG channels are available (as in the WAY-EEG-GAL dataset), more than one muscle could be used to classify motor tasks. Figs.~\ref{fig:comboLH}-~\ref{fig:comboSS} prove that increasing the number of considered muscles leads on average to a better accuracy.

Here, given a specific number of muscles to combine, we randomly selected which muscles to use, we ran the classification with those muscles and then we took the mean and the standard deviation of the set of all accuracy values achievable with the same number of muscles.
Note that to select one muscle among the available five, there are $5$ possibilities; for two and three muscles, $10$ combinations can be extracted; finally, all five muscles are combined to compute the last point of each curve of Figs.~\ref{fig:comboLH}-~\ref{fig:comboSS}.

As a result, the combination of all five available muscles brings the best average accuracy (blue solid line in the figures).
As expected, the standard deviation (blue dashed lines) generally decreases as the number of combined muscles increases.
It is interesting to observe that the variability of the performance in case of surface friction classification is significantly lower than that of the weight classification irrespective of the kernel choice: linear kernel on light/heavy classes provides a similar amount of variability, even though the figure is not reported here for space constraints.
It is also worth mentioning that, by considering the best case (red line) only, the optimal performance is achieved already with three muscles: adding more signals does not further improve the maximum achievable accuracy (saturation effect), but can increase the robustness of the classification algorithm, decreasing the variability of the outcomes.

The value of $0.9538$ is the highest classification performance obtained in the light/heavy case with $4$~s segment duration by combining three specific muscles (best case). At the same time, if the best combination of muscles is not known, the same Figs.~\ref{fig:comboLH}(c) and ~\ref{fig:comboSS}(c) shows that using all muscles with $4$~s segments can yield an accuracy of $0.9421$.

Especially, including more than one muscle obtains a significant advantage in the classification of different friction surfaces: indeed, the best single-muscle accuracy is $0.69$, whereas combining two muscles leads to a value of $0.8$.

\begin{figure*}[htbp]
	\centering
	\resizebox {1\textwidth}{!} {
	\centering
	\captionsetup{justification=centering}
	\subfigure[]
	{
%
\begin{tikzpicture}

\begin{axis}[%
width=4.446in,
height=3.332in,
at={(0.833in,0.657in)},
scale only axis,
area style,
separate axis lines,
every outer x axis line/.append style={black},
every x tick label/.append style={font=\color{black}},
every x tick/.append style={black},
xmin=0,
xmax=6,
xlabel style={font=\bfseries},
xlabel={No. Muscles},
every outer y axis line/.append style={black},
every y tick label/.append style={font=\color{black}},
every y tick/.append style={black},
ymin=0.5,
ymax=1.1,
ylabel style={font=\bfseries},
ylabel={Accuracy},
axis background/.style={fill=white},
xmajorgrids,
ymajorgrids,
legend style={at={(0.97,0.03)}, anchor=south east, legend cell align=left, align=left, draw=black}
]
\addplot[fill=none, draw=none, forget plot] table[row sep=crcr]{%
1	0.617190182764357\\
2	0.704847726050497\\
3	0.771992812578671\\
4	0.803139588875077\\
5	0.850296296296298\\
}
\closedcycle;
\addplot[fill=white!70!blue, draw=none, forget plot] table[row sep=crcr]{%
1	0.195930745582395\\
2	0.142526770121229\\
3	0.0929361855422488\\
4	0.0707874889165161\\
5	0\\
}
\closedcycle;
\addplot [color=blue, line width=1.0pt, forget plot]
  table[row sep=crcr]{%
1	0.617190182764357\\
2	0.704847726050497\\
3	0.771992812578671\\
4	0.803139588875077\\
5	0.850296296296298\\
};
\addplot [color=blue, line width=1.0pt, forget plot]
  table[row sep=crcr]{%
1	0.813120928346753\\
2	0.847374496171726\\
3	0.86492899812092\\
4	0.873927077791593\\
5	0.850296296296298\\
};
\addplot [color=red, line width=2.5pt, mark size=4.0pt, mark=asterisk, mark options={solid, red}]
  table[row sep=crcr]{%
1	0.840296296296299\\
2	0.865629629629632\\
3	0.871111111111114\\
4	0.870444444444447\\
5	0.850296296296298\\
};

\addplot [color=blue, line width=2.5pt, mark=o, mark options={solid, blue}]
  table[row sep=crcr]{%
1	0.715155555555555\\
2	0.776111111111112\\
3	0.818460905349796\\
4	0.838533333333335\\
5	0.850296296296298\\
};

\addplot [color=blue, dashed, line width=2.0pt, name path=B]
  table[row sep=crcr]{%
1	0.813120928346753\\
2	0.847374496171726\\
3	0.86492899812092\\
4	0.873927077791593\\
5	0.850296296296298\\
};

\addplot [color=blue, dashed, line width=2.0pt, forget plot, name path=A]
  table[row sep=crcr]{%
1	0.617190182764357\\
2	0.704847726050497\\
3	0.771992812578671\\
4	0.803139588875077\\
5	0.850296296296298\\
};

\addplot [
        thick,
        color=blue,
        fill=blue, 
        fill opacity=0.4
    ]
    fill between[
        of=A and B,
        soft clip={domain=0:1},
    ];   
\end{axis}
\end{tikzpicture}%
}
	\hspace{5mm}
	\subfigure[]
	{
%
\begin{tikzpicture}

\begin{axis}[%
width=4.446in,
height=3.332in,
at={(0.833in,0.657in)},
scale only axis,
area style,
separate axis lines,
every outer x axis line/.append style={black},
every x tick label/.append style={font=\color{black}},
every x tick/.append style={black},
xmin=0,
xmax=6,
xlabel style={font=\bfseries},
xlabel={No. Muscles},
every outer y axis line/.append style={black},
every y tick label/.append style={font=\color{black}},
every y tick/.append style={black},
ymin=0.5,
ymax=1.1,
ylabel style={font=\bfseries},
ylabel={Accuracy},
axis background/.style={fill=white},
xmajorgrids,
ymajorgrids,
legend style={at={(0.97,0.03)}, anchor=south east, legend cell align=left, align=left, draw=black}
]
\addplot[fill=none, draw=none, forget plot] table[row sep=crcr]{%
1	0.64041788320893\\
2	0.737850097211756\\
3	0.821509704920601\\
4	0.837296743776039\\
5	0.887851851851856\\
}
\closedcycle;
\addplot[fill=white!70!blue, draw=none, forget plot] table[row sep=crcr]{%
1	0.217623492841399\\
2	0.174759064835752\\
3	0.0944126889242356\\
4	0.0810509568923721\\
5	0\\
}
\closedcycle;
\addplot [color=blue, line width=1.0pt, forget plot]
  table[row sep=crcr]{%
1	0.64041788320893\\
2	0.737850097211756\\
3	0.821509704920601\\
4	0.837296743776039\\
5	0.887851851851856\\
};
\addplot [color=blue, line width=1.0pt, forget plot]
  table[row sep=crcr]{%
1	0.858041376050329\\
2	0.912609162047507\\
3	0.915922393844837\\
4	0.918347700668411\\
5	0.887851851851856\\
};
\addplot [color=red, line width=2.5pt, mark size=4.0pt, mark=asterisk, mark options={solid, red}]
  table[row sep=crcr]{%
1	0.896888888888893\\
2	0.927481481481486\\
3	0.916074074074078\\
4	0.907703703703708\\
5	0.887851851851856\\
};

\addplot [color=blue, line width=2.5pt, mark=o, mark options={solid, blue}]
  table[row sep=crcr]{%
1	0.74922962962963\\
2	0.825229629629632\\
3	0.868716049382719\\
4	0.877822222222225\\
5	0.887851851851856\\
};

\addplot [color=blue, dashed, line width=2.0pt, name path=B]
  table[row sep=crcr]{%
1	0.858041376050329\\
2	0.912609162047507\\
3	0.915922393844837\\
4	0.918347700668411\\
5	0.887851851851856\\
};

\addplot [color=blue, dashed, line width=2.0pt, forget plot, name path=A]
  table[row sep=crcr]{%
1	0.64041788320893\\
2	0.737850097211756\\
3	0.821509704920601\\
4	0.837296743776039\\
5	0.887851851851856\\
};

\addplot [
        thick,
        color=blue,
        fill=blue, 
        fill opacity=0.4
    ]
    fill between[
        of=A and B,
        soft clip={domain=0:1},
    ];

\end{axis}
\end{tikzpicture}%
}
	\hspace{5mm}
	\subfigure[]
	{
%
\begin{tikzpicture}

\begin{axis}[%
width=4.446in,
height=3.332in,
at={(0.833in,0.657in)},
scale only axis,
area style,
separate axis lines,
every outer x axis line/.append style={black},
every x tick label/.append style={font=\color{black}},
every x tick/.append style={black},
xmin=0,
xmax=6,
xlabel style={font=\bfseries},
xlabel={No. Muscles},
every outer y axis line/.append style={black},
every y tick label/.append style={font=\color{black}},
every y tick/.append style={black},
ymin=0.5,
ymax=1.1,
ylabel style={font=\bfseries},
ylabel={Accuracy},
axis background/.style={fill=white},
xmajorgrids,
ymajorgrids,
legend style={at={(0.97,0.03)}, anchor=south east, legend cell align=left, align=left, draw=black}
]
\addplot[fill=none, draw=none, forget plot] table[row sep=crcr]{%
1	0.633343770278813\\
2	0.824801126526454\\
3	0.846206987924398\\
4	0.872682391576004\\
5	0.942074074074079\\
}
\closedcycle;
\addplot[fill=white!70!blue, draw=none, forget plot] table[row sep=crcr]{%
1	0.274912459442376\\
2	0.117294866288661\\
3	0.107200838966026\\
4	0.0952574390702237\\
5	0\\
}
\closedcycle;
\addplot [color=blue, line width=1.0pt, forget plot]
  table[row sep=crcr]{%
1	0.633343770278813\\
2	0.824801126526454\\
3	0.846206987924398\\
4	0.872682391576004\\
5	0.942074074074079\\
};
\addplot [color=blue, line width=1.0pt, forget plot]
  table[row sep=crcr]{%
1	0.908256229721189\\
2	0.942095992815116\\
3	0.953407826890424\\
4	0.967939830646227\\
5	0.942074074074079\\
};
\addplot [color=red, line width=2.5pt, mark size=4.0pt, mark=asterisk, mark options={solid, red}]
  table[row sep=crcr]{%
1	0.93718518518519\\
2	0.946962962962969\\
3	0.953777777777783\\
4	0.948740740740746\\
5	0.942074074074079\\
};

\addplot [color=blue, line width=2.5pt, mark=o, mark options={solid, blue}]
  table[row sep=crcr]{%
1	0.770800000000001\\
2	0.883448559670785\\
3	0.899807407407411\\
4	0.920311111111115\\
5	0.942074074074079\\
};

\addplot [color=blue, dashed, line width=2.0pt, name path=B]
  table[row sep=crcr]{%
1	0.908256229721189\\
2	0.942095992815116\\
3	0.953407826890424\\
4	0.967939830646227\\
5	0.942074074074079\\
};

\addplot [color=blue, dashed, line width=2.0pt, forget plot, name path=A]
  table[row sep=crcr]{%
1	0.633343770278813\\
2	0.824801126526454\\
3	0.846206987924398\\
4	0.872682391576004\\
5	0.942074074074079\\
};

\addplot [
        thick,
        color=blue,
        fill=blue, 
        fill opacity=0.4
    ]
    fill between[
        of=A and B,
        soft clip={domain=0:1},
    ];

\end{axis}
\end{tikzpicture}%
}
	}
  \caption{Accuracy vs no. of muscles used for the classification of light/heavy objects using (a) $1$~s segments, (b) $2$~s segments or (c) $4$~s segments. Mean (blue solid line) and standard deviation (blue dashed lines) of the accuracy (among all possible combinations) together with the best case (red line) are represented.}
\label{fig:comboLH}
\end{figure*}
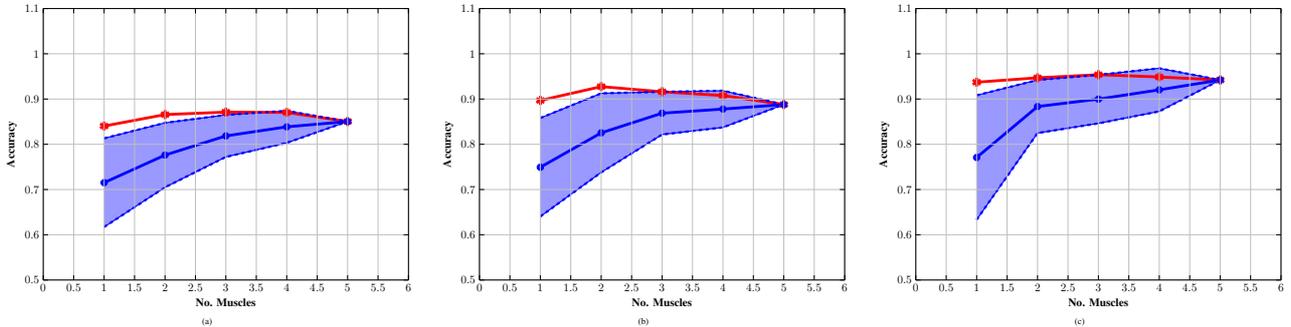

\section{Discussion and Conclusions} \label{sec:discussion}

This work presents an innovative application of CMC for the classification of different motor tasks, i.e., grasping of different kinds of objects.

\subsection*{On CMC ability to classify haptics}
CMC is able to reliably classify objects with different weights and surface frictions. As far as we know, the ability to accurately recognize haptic features of an object (e.g., sandpaper vs silk surfaces), is a completely new result in neuroscience.

We show that after training SVM on CMC samples (eight samples, one per each frequency band of interest) with a linear or RBF kernel, a very high accuracy can be achieved in case of both light/heavy objects (over $0.9$) and sandpaper/silk cover(over $0.8$).

Accuracy is further improved if larger data segments, i.e., longer observation periods, are considered. This is especially evident for light/heavy classification and could be expected as longer segments in time provide higher frequency resolution and then more accurate evaluations of the CMC in the frequency domain.

Another interesting aspect is the use of different kernels for different classes of data: unexpectedly, the linear kernel provides higher performance for light/heavy classification, while RBF turns out to be better for sandpaper/silk surfaces. This suggests that the SVM kernel can be adapted to different kinds of data, which might be considered as a key aspect to further investigate.

Finally, it can also be noted that the variability in case of different friction surfaces is significantly lower than in the case of different weights.
This might be related to the kind of information carried by muscles in relation to the specific motor task: it can be suggested that for the correct classification (accuracy above $0.8$) of haptic features of the object, i.e. friction surface, all considered muscles bring useful and equally important information, so that their joint analysis is recommended. 
On the contrary, large standard deviations found, e.g., for the weight classification can be addressed to the combination of more heterogeneous information that possibly includes \emph{side} contributions, i.e., not relevant to the classification.

Some limitations still affect the present work: above all, a limited dataset with few specific motor tasks has been considered; however, the same analysis could be easily extended in the future to other datasets and other kinds of motor tasks.

Moreover, it is well-known that specific frequency bands can contribute to CMC more than others: e.g., $\beta$ band strongly determines the brain-muscles communications during static motor output (i.e., isometric contractions), while $\gamma$ band is more related to dynamically changing movements\cite{gwin}. Besides, the frequency bands in which CMC takes values strongly depend on the amount of applied force during the motor task~\cite{lattari}: CMC contains large $\beta$ band components if weak contractions are produced, whereas $\gamma$ band is larger in case of strong contractions.
Therefore, one can expect that the classification accuracy changes if only specific frequency bands are used. Such kinds of considerations could be tested in a future classification analysis to discern whether accuracy in classification is provided by either the broad CMC frequency spectrum (as used here) or few specific frequency bands.
This can further provide valuable insights on the mechanisms of synchronization between brain and muscles during movement.

\subsection*{Impact and perspectives}
The impact of this work mainly consists of the insights provided by CMC on the synchronized brain-muscles activation patterns during the accomplishment of different kinds of motor tasks: specifically, the analysis presented so far gives precise information about the amount of relevant brain-muscles synchronization during common motor activities (like holding different kinds of objects). This can be interpreted as an indication of the synergy existing between the brain and the muscles to perform a specific activity and might support the hypothesis of a predominant centralized control of motor output (in opposition to the hypothesis of a distributed control of \emph{motor synergies})~\cite{DAvella}. 

Besides, it is worth mentioning that movement has been previously studied with different approaches in different scientific communities: in neuroscience, the complex and computationally expensive tool of motor synergies was used to successfully reconstruct several kinds of movements, from rats to primates. CMC was also developed in the same field and mainly used to investigate motor tasks where different levels of force were exerted and when contraction was dynamically changed\cite{mima}. 
On the other hand, the most common approach in the engineering and robotics communities made use of inertial sensors (gyroscopes and accelerometers) to track several limb joints during the movement and, throughout a reconstruction of all body segments, the classification was performed~\cite{INERTIAL}. 

With this work we suggest the importance to make a bridge between the two communities and in particular to test the fully innovative use of CMC for motion analysis and activity recognition in broader terms.
With this same perspective, it might be worth to plan a new measurement campaign to simultaneously record both inertial measures and electrophysiological data (EEG and EMG), in order to compare the different classification approaches and their effectiveness in motion activity recognition, from rough and easy movements to the finest.

In fact, it might be expected that the joint perspective between the two communities can bring improvements in the classification of haptics, providing brand-new outcomes that could be highly beneficial for rehabilitation from motor diseases.

\end{document}